# A DECISION-MAKING SUPPORT SYSTEM BASED ON KNOW-HOW

Kryssanov, V.V., Abramov, V.A., Fukuda[*], Y., Konishi[*], K.

IACP, 5, Radio St., Vladivostok, 690041 Russia
[*] TRI of JSPMI, 1-1-12 Hachiman-cho, Higashikurume-shi, Tokyo, 203 Japan
kryssano@jspmi.or.jp

**ABSTRACT**

The research results described are concerned with:
- developing a domain modeling method and tools to provide the design and implementation of decision-making support systems for computer integrated manufacturing;
- building a decision-making support system based on know-how and its software environment.

The research is funded by NEDO, Japan.

## 1. INTRODUCTION

Today, one of the important problems of Computer Integrated Manufacturing (CIM) is to utilize the human experience of making decisions in a variety of professional tasks accumulated by leading manufacturers over the world, and propagate this experience by developing appropriate computer-aided support technologies. From a computer science perspective, it is crucial to create a strong theoretical foundation and build tools to acquire, analyze and synthesize the information of decision-making.

Extensive researchers' efforts have been made on understanding and formalizing the activity of decision-making in the manufacturing domain since the 80s. As a result, a number of systems qualified to support the activity have been developed. These systems can be naturally classified into three categories: information retrieval systems, decision support systems and expert systems. As no clear distinction between these categories can be found in the literature, we suggest that a distinction should be made based on the notions that:
- an information retrieval system is a computer-based system to capture, manipulate, retrieve and transmit organized data necessary to solve a professional task according to detailed transactions defined by a user;
- a decision support system is a knowledge-based information system to capture, handle and analyze information which affects or is intended to affect decision-making performed by people in the scope of a professional task appointed by a user;
- an expert system is a knowledge-based system to be used instead of or together with a human operator to make decisions in the framework of a professional task with explanations for users.

This is an informal classification of the systems dealing with decision-making in manufacturing. However, it may be helpful to realize what has already been done and how it corresponds with present and future trends of the manufacture infrastructure.

It should be noted first, that information retrieval systems are currently not discussed in the literature as standalone programs but as a part of specialized CIM applications: CAD, CAM, CAPP, etc. There are many reasons for this, and the dissimilarity of manufacturing terminology can be pointed to as one. Another reason is existing retrieval systems cannot provide the freedom in formulating transactions necessary for CIM.

Knowledge-based systems have been developed for manufacturing for more than twenty years. Despite many limitations in the points of applicability, integration and maintenance of the systems, research in this area is considered now as a core of computer-aided decision-making in CIM. At the same time, expert systems, which make decisions automatically, are not discussed as an inevitable component of the manufacturing environment at present (Wada, 1996). So, hereafter we will concentrate on decision support systems as well as interactive expert systems, referring to them as DSS.

A further distinction needs to be drawn between stages of maturation of the support technologies for decision-making. We will argue here that any idea concerning the technologies, to be successfully put into practice, should have an adequate embodiment on each of these three levels of problem awareness: 1) formal specification, 2) design, 3) implementation and maintenance.

To apply a formal specification technique to decision-making means to use a formal method having a rigorous mathematical basis and capable of modeling the domain in general, specifying the domain professional tasks and representing the domain knowledge as well. The formal method is to provide decision support applications design with the highest integrity, reliability and compatibility.

On the stage of design, it is necessary to explicate the mathematical and program content of the technology.

Taking the characteristics, which a user expects to witness in dealing with a DSS for a basis, functionality and behavior of the DSS and its software environment are decided.

Implementation and maintenance of a decision-making support technology assumes implementation of a DSS as well as the systems to ensure the functions of updating, verification, validation and integration with external applications of the DSS which are necessary.

An overwhelming majority of research reported in the literature has been on the detailed design of decision support systems, whereas formal specification of the problem has still been a difficult and expensive task. Few projects have been brought to the stage of implementation and maintenance, and no attempts have been published to give a description of a support technology for decision-making in CIM for all of the above levels of problem awareness.

The focus of this paper is on a scientific method for developing and maintaining Decision Support Systems as essential components of CIM. In the following section, we formulate the scope and necessary assumptions of the method. In Section 3, an approach to modeling the domain and specifying the domain tasks is outlined. Then we discuss some aspects of the information representation technique. A view of the decision-making process and a software environment to support decision-making technologies are presented in Section 5. Next, the suggested method is extended to the case of utilizing empirical know-how. Section 7 gives an account of an experiment resulting in a DSS and its software environment. Finally, in Section 8, we discuss related work and draw some conclusions from our experiences.

## 2. STARTING POINTS

Despite the fact that no universal approach to decision-making in CIM exists at present, a number of instances of success using various knowledge-based and algorithm techniques of making decisions for domain tasks can be found in the literature. As specification of the input and output of the task to be solved with a particular method is still necessary to develop a DSS, these instances might be considered as the knowledge (or metaknowledge) of how it is possible to get desired information under the conditions. Such interpretation makes the knowledge-based approach more fundamental. However, the problem is how to integrate the methods, qualitative as well as quantitative, within one DSS.

We guess, that a basis for such integration can be a mapping of the domain professional activity onto a unified information structure, i.e., a domain model.

We will consider that the domain is characterized by the professional activity that consists in solving different tasks (or else we will have to think about a DSS owning encyclopedic knowledge). Solving a task takes place in the domain reality, requires professional knowledge, and affects a set of domain objects, both physical and conceptual. We will assume that the domain reality can be thought of as a potentially infinite set of separated situations, where each situation corresponds to a DSS run made to solve a professional task. Rejecting this hypothesis leads us to the conception of real-time systems that is out of the scope of this paper. The situations are not linked to one another. This assertion causes the discrete updating of the knowledge for a DSS. Each of the situations is defined on a finite set of domain objects, that means time and space frames of a situation as well as human ability to recognize a finite set of objects within a professional task. The professional knowledge is how people use the domain objects to solve the task when the information about the objects is used as the input, output and intermediate data. This knowledge, the input data and the output data for every task can be represented verbally. The latest assumption is always true for input and output data (if not, decision-making support becomes meaningless), but it does not interdict implicit or tacit knowledge to be used.

## 3. THE FORMALISM

Let us introduce the necessary notions.

Definition 1. An $S$-sorted set $A$ is a family of sets $A_s$, one for each sort $s \in S$: $A = \{A_s \mid s \in S\}$. ∎

Definition 2. A many-sorted signature is a couple $\langle S, \Sigma \rangle$, where $S$ is the sort set, $\Sigma$ is an $S^* \times S$-sorted family $\{\Sigma_{a,s} \mid a \in S^*, s \in S\}$ of symbols, and every symbol $\sigma \in \Sigma_{a,s}$ has arity $a$ and sort $s$. ∎

Definition 3. An algebra $\mathcal{A}$ of a many-sorted signature $\langle S, \Sigma \rangle$ is a finite family $A = \{A_s \mid s \in S\}$ of sets called the carriers of $\mathcal{A}$, conjointly with a mapping $A_\sigma : A_a \to A_s$ for each $\sigma \in \Sigma_{a,s}$, where $A_a = A_{s1} \times \cdots \times A_{sn}$ when $a = s1 \cdots sn$. ∎

Let $_X\Sigma$ be a signature $\Sigma$ with a finite set of variables $X$ such that $\Sigma = \bigcup_{i=0}^{n} \Sigma^i$, $n > 0$, $\Sigma^i \bigcap_{i \neq j} \Sigma^j = \varnothing$, and $\Sigma^i$ is a many-sorted signature with the sort set $S = \{O, F, P\}$, here $O, F$ and $P$ are respectively objective, functional and predicate sorts; $X = \{X_s \mid s \in S\}$ is disjointed from $\Sigma$.

Now we introduce a declarative model for representing the domain as the following.

Definition 4. A domain model $M^n$ of n-th order with the signature $_X\Sigma$ is a tuple $M^n = \langle \Lambda, \Phi \rangle$ such that

$\Lambda = \{\mathcal{A}^0, \mathcal{A}^2, \cdots, \mathcal{A}^n\}$ for $n > 1$, and $\Lambda = \{\mathcal{A}^0\}$ for $n = 1$; $\mathcal{A}^i$ is an algebra of the signature $\Sigma^i$, $i = 0, 2, 3, \cdots, n$. The carriers $A^i$ of the algebra $\mathcal{A}^i$, $i > 1$, follow the condition: $\Sigma_O^0 \subset A^2$, $\Sigma^{i-1} \subset A^i$ and $A^i \subset A^{i+1}$ when $i = 2, \cdots, n$. There are partial mappings $A_\sigma : B \to C$ when $\sigma \in \Sigma_F$, $A_\sigma : B \to D$ when $\sigma \in \Sigma_P$, and

$A_\sigma : D \to C$ when $\sigma \in \Sigma_O$. Here $B$ is a Cartesian multiplication of nonempty subsets of $A^i$, $C$ is a nonempty subset of $A^i$, $D$ is a one-point set.

$\Phi$ is a finite set of quantifier-free logical formulas involving variables of $X$. These formulas are written in a predicate calculus language defined on $\Sigma$.[†] ∎

The predicate calculus language is of the languages based on higher order logics (in general case) where terms and formulas can be in the form: $x^{(k)}(t_1, \cdots, t_m)$, here $x^{(k)} \in X$ is a variable, $t_1, \cdots, t_m$ are terms, $1 \leq k \leq n$ is the order of the variable.

The range of values of $x^{(k)}$ is a subset of $A^0$ if $k=1$, and a subset of $A^k$ for $k>1$. We will call an $S$-sorted function $R^i = \left\{ R_s : Z_s^i \to A_s^i \,\middle|\, Z_s^i \subset X_s, s \in S \right\}$, $i>0$, as an assignment of values in $A^i$ to variables in a variable set $X$.

<u>Definition 5.</u> An algebra $\mathcal{A}^1$ of the signature $\Sigma^1$ such that $\Sigma_O^0 \subset A^1$ and $A^1 \subset A^2$ is a solution for the model $\mathrm{M}^n$ iff every formula entering $\Phi$ is true for all the possible $R^i$, $i=1,\cdots,n$. ∎

It is important to note that the symbols of $\Sigma^1$ will be interpreted as unknowns for the model.

We will specify the input data and conditions of a domain professional task in terms of $\Sigma^0 \bigcup \Sigma^1$ (including at least one symbol of $\Sigma^1$), the output data – in terms of $\Sigma^1$, and the optimization criterion for the task – as a predicate defined on $\mathcal{A}^1$.

More details of this formalism can be found in Kleshchev, et al., (1996).

## 4. INFORMATION REPRESENTATION

To consider computer-aided decision-making support based on the domain model, we need to formulate the main principles for mapping the domain information onto the logical structure described above.

Let us define an abstract set of domain objects such that: a) all of the domain objects in the set, i.e., the instances have the same identifying attributes, b) all the instances are subject to and conform to the same rules. A representation of common properties of elements of such a set will be called a domain scale. Naturally, decision-making uses "values on the scales", i.e., information models of the set elements, instead of domain objects. A set of domain scales retaining all the domain objects forms the domain scale system. Taxonomies, product data representation standards or the like are essentials of the domain scale system.

We will regard the algebra $\mathcal{A}^0$ as a unified representation of the domain scale system.

It can be postulated that the information of the domain reality has a stable structure. A model of this structure is a domain ontology. We will argue that an ontology can be created with a collection of content-specified knowledge representation primitives. In general, this collection consists of object constants, relations and classes. Situation semantics (type of action, state, relationships, which are described in the situation) are organized around an object constant within the situation that is a dedicated domain object. A relation is a set of $m$-tuples, $m>1$, where each of the tuples links a finite set of domain objects into a sequence and represents a relationship among objects arising while solving a task. There are special types of relations called functional relations, where the last item of an $m$-tuple is the value of the function on the first $m-1$ items of the tuple. A class is a unary relation, i.e., a set of tuples of length one.

We will consider the symbols of $\Sigma^1$ as designations of the primitives, which make up an ontology.

Let an abstraction be a projection reducing granularity of the domain information. It can be thought of that the algebra $\mathcal{A}^i$, $i>0$, represents a set of facts (those are instances of how subject matter experts treat the domain objects during solving professional tasks) mapped onto abstractions of the same level. Then, the granularity is decreased from fine to coarse by increasing $i$.

The set of formulas $\Phi$ is created by consistency constraints on $\mathcal{A}^i$, $i=1\ldots n$, and the formulas determining connections between $\mathcal{A}^i$, $i>1$, and $\mathcal{A}^1$. Making decisions involves resolving formulas of $\Phi$.

Now we define a DSS as a practical accomplishment of the domain model $\mathrm{M}^n$ of the signature $_X\Sigma$ having a finite solution set and a collection of solving methods for specified professional tasks.

Clearly, a model of a signature $_X\dot\Sigma = \Sigma^0 \bigcup \dot\Sigma^1 \bigcup \dot\Sigma^2 \bigcup \cdots \bigcup \dot\Sigma^n$ and a model of a signature $_Y\ddot\Sigma = \Sigma^0 \bigcup \ddot\Sigma^1 \bigcup \ddot\Sigma^2 \bigcup \cdots \bigcup \ddot\Sigma^m$ can have the same solution set. This condition provides the model tractability necessary during the stage of DSS design.

Let $\mathrm{M}^n$ be a model with $\mathcal{P}$ a potentially infinite set of solutions. Then, $\mathrm{M}^n$ might be an information model of the whole domain reality.[‡] We will argue that establishing computer-aided support for decision-making in CIM predestinates building $\mathcal{P}$ such that every task in the domain reality has at least one model in $\mathcal{P}$. Practically, it means that there is continued updating of the domain information held by DSS. This is a reason to develop a fund of domain information (knowledge) as a global model $\mathrm{M}_G^n$ with the solution set $\mathcal{P}$ as a unified core for decision support systems. Then, while the model $\mathrm{M}_G^n$ is modified as necessary, a DSS can utilize a part of the domain information represented in $M_G^n$. The latest possible if $\mathcal{A}^0 \subseteq \mathcal{A}_G^0$ and $\mathcal{A}^1 \subseteq \mathcal{A}_G^1$, where $\mathcal{A}^1, \mathcal{A}^0$ are of the model realized in the DSS, and $\mathcal{A}_G^1, \mathcal{A}_G^0$ are of $M_G^n$.

---

[†] This stipulates applying predicate calculus logic rather than a programming technique.

[‡] However, the reader should be aware that such a model could embody no models or one or many models for a task.

Recently, there is the researchers' belief that capturing manufacturing ontologies has great implications for developing CIM information-based applications (Benjamin, et al, 1995). Because people, as a rule, have great difficulty articulating the knowledge associated with the subconscious and skill-based activity, elicitation of ontologies (necessarily included with such knowledge) appears to be a difficult problem. In the framework of our approach, the implicit knowledge would be represented by a wide range of domain tasks with their associated responses - the domain facts. These facts are appended to $M_G^n$. And the implicit component of $\mathcal{P}$ would be acquired through mapping $\mathcal{A}_G^i$ ($i > 1$) onto $\mathcal{A}_G^1$ accordingly as shown with the formulas of $\Phi_G$. This method of knowledge acquisition is a kind of "training on instances" that has several drawbacks. The main predicament is the necessity to analyze a huge amount of facts that vary on many dimensions. We suppose, that establishing an associated domain information space through the Internet would make this problem manageable.

## 5. DECISION-MAKING IN CIM

Figure 1 presents a conceptual diagram of the decision-making process in the domain. As it is demonstrated in the figure, decision-making is initiated by an external inquirer (it can be a user or a system) and begun from gathering information to specify a problem (task). Then, associated information is recognized (this information can be used to refine the problem), and relevant (local) reality models are selected. The models are evolved, and possible alternatives (problem solutions) are evaluated. Next, the best solutions are chosen for the final decision. After the decision is made, the new reality is considered and, if necessary, the process is repeated under new conditions.

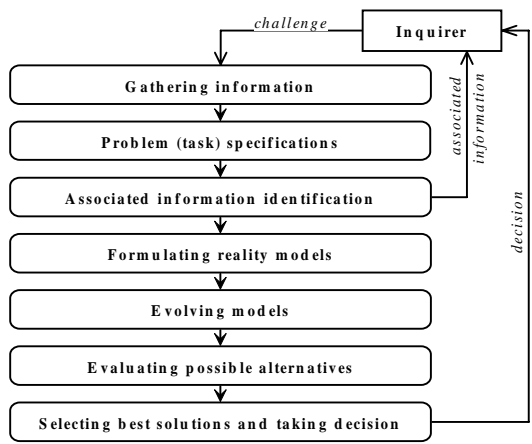

**Fig. 1. Decision-making process.**

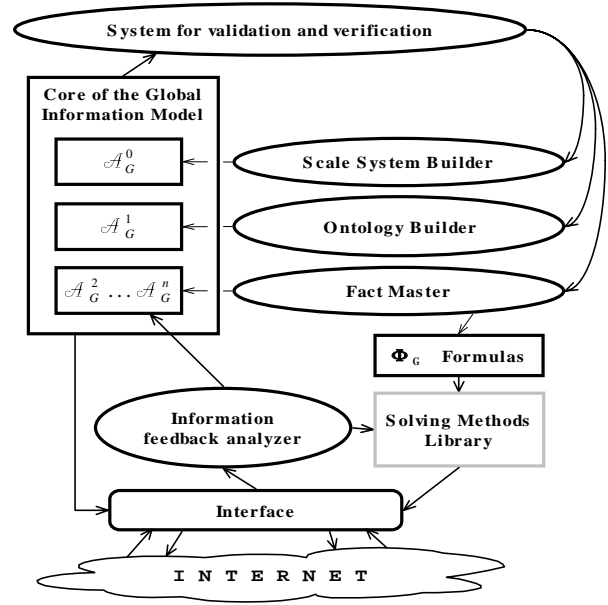

**Fig. 2. Support environment for decision-making.**

Decision-making in CIM has the same structure, however it is necessary to consider communication aspects to provide group collaboration support. The latest means, the usage of structured group decision techniques (Forgionne, 1991) via an information layer to manage shared information processing is one feasible way.

We define a decision-making technology in CIM as a technique of applying DSS to make decisions for a special problem. So, as it was formulated in Section 4, to establish computer-aided support for decision-making technologies is, first, to create a unified information space and, second, to build the domain reality model.

Several interactive computer information tools should be developed to deliver complete support for communication and decision-making. Being linked through the Internet, they form an application environment for DSS. Figure 2 shows the main components in the environment based on the global domain model, what the relationships are among them, and how they handle the domain information. The following systems are needed:

- *Scale System Builder:* a tool to construct and update the domain scale system as a set of abstract data types.
- *Ontology Builder:* a CASE system to arrange models of domain ontologies. There are three predefined symbol groups: object constants, relations and classes (see Section 4). Symbols of the groups are combined in the structure that can be completely represented within the first order logic.
- *Fact Master:* a tool to collect and represent the domain facts. There is a knowledge representation language based on higher order logics.
- *System for validation and verification:* a tool to check pertinency of the domain facts to the reality model and the scale system. If there is incompleteness of $\mathcal{A}^0$ or $\Sigma^1$, or inconsistency of $\mathcal{A}^1$, the system delivers the related

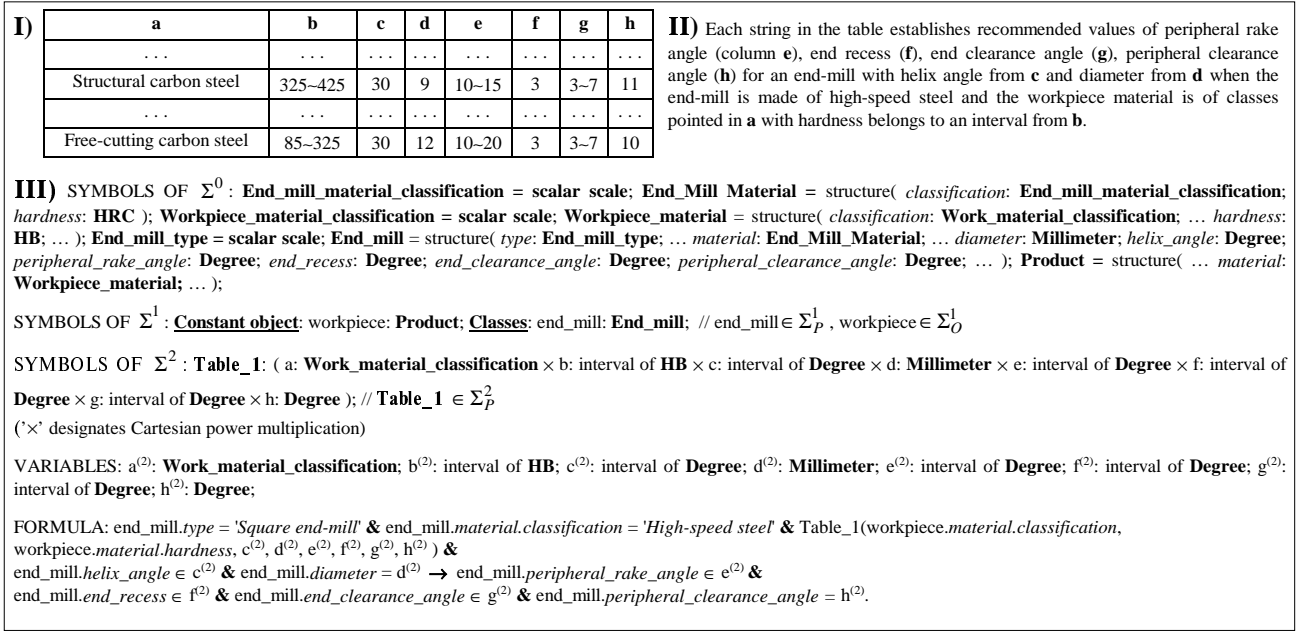

**Fig. 3. An example of know-how formalization.**

information that may be used by people to modify the domain model core.

- *Interface:* an automatic system to provide access to the global information model for CIM applications. Necessary portability and interoperability on the level of information representation models can be enabled with semantic translation by articulation axioms (Collet, et al., 1991).
- *Information feedback analyzer:* a tool to decompose information of practical instances of taking decisions represented in terms of $\Sigma^0 \cup \Sigma^1$ into declarative components (new facts to be added to the domain model) and procedural components (new samples of problem solving methods to be added to the Solving Methods Library).

## 6. KNOW-HOW UTILIZATION

It is a well-known fact, that know-how plays a significant role during decision-making by people in actual manufacturing. Generally, the knowledge of know-how can be classified into the conjectural knowledge (having mostly a procedural nature) and experimental knowledge (having largely a declarative character). The former is usually a generalization of professional skill while the latter emanates from the results of practical experiments. There are distinct difficulties in utilizing these kinds of knowledge for DSS: conjectural knowledge is intricately acquired and represented whereas empirical knowledge is hardly used. A key to solving the first problem would be gathering a representative collection of instances of treatment of a problem by subject matter experts, and then mapping the information of this collection onto the domain reality model as it was drawn in the previous sections. The experimental knowledge usually comes to be known in a form already formalized - as tables or graphs, - which is rather easy to represent in DSS. However, extremely large amounts of such know-how, their peculiar nature, and high granularity make the experimental knowledge difficult to utilize widely for application. A way to manage this problem would be to accumulate such know-how by representing them in algebras $\mathcal{A}_G^i$, $i > 1$, and formulas of $\Phi_G$.

A variety of know-how has been collected from Japanese manufacturers in this study. An example of that material is partially reproduced in Figure 3. The knowledge of the example may be accessed while selecting tools in process planning. Part (I) of the figure shows a fragment of the table containing the information about end mill edge angles recommended to prolong tool life. Part (II) summarizes how to use the information of the table, and part (III) systematically introduces elements of model $M^n$ which would be to add the information of the example to the model (simplifying the rather lengthy formal representation).

## 7. RESULTS

A working approximation of the application environment described in Section 5 has been developed in the research. Prototypes of a Scale System Builder and Ontology Builder have been implemented as one software tool.[§] 47 examples of know-how treated of end milling have been analyzed, and a DSS based on this know-how has been developed using the tool (see Figure 4). A reality model for the task of calculating the machining parameters has been realized in the DSS. The knowledge base of the system contains logical formulas including ten formulas with higher order predicates. Taking the characteristics of the product material and the end-mill as the input data, the system is able to get a solution for two optimization criteria: tool life or machining operation time. Forward chaining is used in the model-based reasoning. A function of explanation (textual and graphic illustrations) is provided by the system. Updating and enlarging the fact (know-how) base of the system are accessible through the software tool. An example of a DSS run is shown in Figure 5. All the above programs have been implemented on IBM PC platform in Borland Pascal 7.0 .

---

[§] In our current implementation, networked communication is not provided.

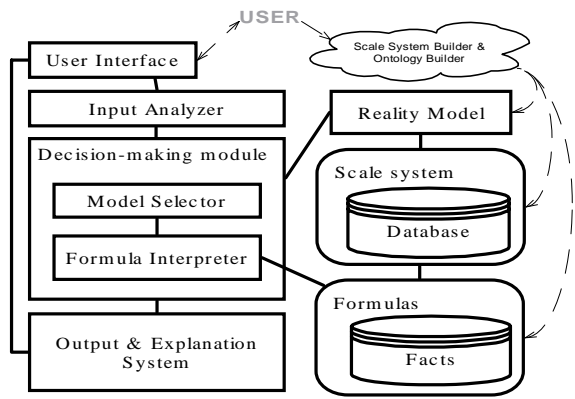

**Fig. 4. Modular structure of the DSS.**

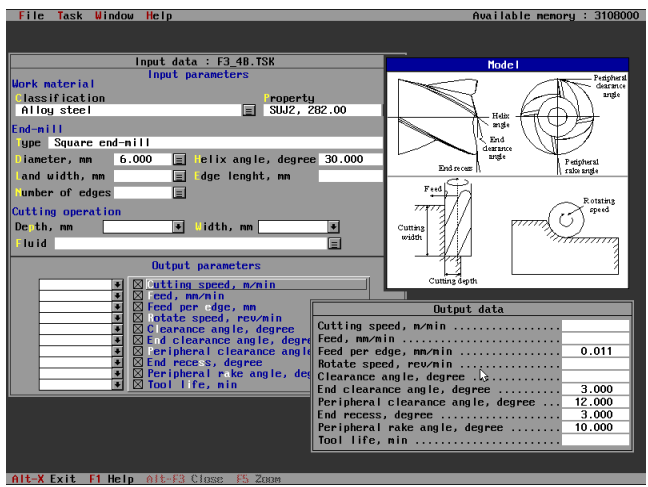

**Fig. 5. A sample run of the system.**

## 8. RELATED WORK AND CONCLUSIONS

Recently, research world-wide has begun on information processing in CIM. (Koonce, et al., 1996) discussed the integration aspects for manufacturing applications and provided a view of the architecture for the integrated manufacturing systems based on the unified data meta-model. In Aguiar and Weston, (1995) a general model-based approach to computer-aided support of the life cycle of integrated manufacturing enterprises was introduced. Demands on the next generation of manufacturing resource data and info-systems were identified by Kjellberg and Bohlin, (1996). Despite much research on the subject being published, almost all of it, including the works mentioned above, have been made on the conceptual level rather than the pragmatic one.

In this paper we have introduced a model-based technique for developing and maintaining Decision Support Systems in CIM and illustrated the approach with an example from our work on the problem of utilizing empirical know-how. A few conclusions can be drawn from our experiences:

- The development of formal methods is an important issue in software engineering technologies for manufacturing. Applying an adequate formal specification technique much facilitates creating decision support technologies in CIM.
- Building a unified domain model accessible through the Internet is a vital task of CIM. Such a model can serve as a core to gather, analyze and synthesize the information for DSS.
- Founding CIM ontologies can assist utilizing know-how for computer-aided decision-making. Capturing the ontologies can be smoothly organized while an associated information space is established in the domain.

One more outcome of the study is that the progress of CIM application systems requires the creation of a domain knowledge (information) fund rather than mere CIM taxonomies and ontologies. A possible way to ensure representation of multiple domain knowledge is the use of a domain-oriented predicate calculus language based on higher order logics.

## REFERENCES

Aguiar, M.W.C., and Weston, R.H. (1995). A model driven approach to enterprise integration. *Int. J. Computer Integrated Manufacturing*, Vol. 8, No. 3, pp. 210-224.

Collet, C., Huhns, M.N., Shen, W.-M. (1991). Resource Integration Using a Large Knowledge Base in Carnot. *IEEE Computer*, Vol.24, No. 12, pp. 55-62.

Forgionne, G.A., (1991). Decision Technology Systems: A Vehicle To Consolidate Decision Making Support. *Information Processing & Management*, Vol. 27, No. 6, pp. 679-697.

Kjellberg, T., Bohlin, M. (1996). Design of a Manufacturing Resource Information System. *CIRP Annals*, Vol. 45/1, pp. 149-152.

Kleshchev, A.S., Kryssanov, V.V., Fukuda, Y., Konishi K. (1996). CAPP Expert System Building Technology: Logical Approach. In: *Proceedings of The 6$^{th}$ IFIP TC5/WG5.7 International Conference APMS'96*, (N.Okino, H.Tamura and S.Fujii (Ed.)), pp. 115-120. IFIP, Kyoto.

Koonce, D.A., Judd, R.P., and Parks, C.M. (1996). Manufacturing systems engineering and design: an intelligent, multi-model, integration architecture. *Int. J. Computer Integrated Manufacturing*, Vol. 9, No. 6, pp. 443-453.

Wada, R. (1996). A new concept for next-generation machine tools. In: *Proceedings of The 7$^{th}$ International Machine Tool Engineers Conference*, pp. 173-189. JMTBA, Tokyo.